\documentclass{aa}

\usepackage{graphicx}
\usepackage{natbib}
\usepackage{color}
\usepackage{colortbl}
\usepackage{hyperref}
\usepackage{amsmath}

\hypersetup{colorlinks=true,allcolors=[rgb]{0,0,0.8}}
\definecolor{grijs}{gray}{0.40}
\definecolor{lichtgrijs}{gray}{0.90}

\makeatletter
\renewcommand*\aa@pageof{, page \thepage{} of \pageref*{LastPage}}

\usepackage{txfonts}
\begin{document}

   \title{Tidally Heated Exomoons around $\epsilon$ Eridani b: Observability and prospects for characterization }


  \author{E. Kleisioti
         \inst{1}\fnmsep
         \inst{2}
          \and
          D. Dirkx\inst{2}
          \and
          M. Rovira-Navarro 
        \fnmsep
          \inst{3}
          \and
          M. A. Kenworthy
          \inst{1}
          }

   \institute{Leiden Observatory, Leiden University, P.O. Box 9513, 2300 RA Leiden, The Netherlands\\
              \email{kleisioti@mail.strw.leidenuniv.nl}
         \and
             Faculty of Aerospace Engineering, TU Delft, Building 62 Kluyverweg 1, 2629 HS Delft, the Netherlands
         \and
            Lunar and Planetary Laboratory, University of Arizona. Tucson, AZ 85721, USA}

   \date{Received 6 February 2023; accepted 2 May 2023}

  \abstract
   {Exomoons are expected to orbit gas giant exoplanets just as moons orbit solar system planets. Tidal heating is present in solar system satellites and it can heat up their interior depending on their orbital and interior properties.} 
   {We aim to identify a Tidally Heated Exomoon's (THEM) orbital parameter space that would make it observable in infrared wavelengths with MIRI/JWST around $\epsilon$ Eridani b. We study the possible constraints on orbital eccentricity and interior properties that a successful THEM detection in infrared wavelengths can bring. We also investigate what exomoon properties need to be independently known in order to place these constraints.}
   {We use a coupled thermal-tidal model to find stable equilibrium points between the tidally produced heat and heat transported within a moon. For the latter, we consider a spherical and radially symmetric satellite with heat being transported via magma advection in a sub-layer of melt (asthenosphere) and convection in the lower mantle. We incorporate uncertainties in the interior and tidal model parameters to assess the fraction of simulated moons that would be observable with MIRI.}
   {We find that a $2 R_{Io}$ THEM orbiting  $\epsilon$ Eridani b with an eccentricity of 0.02, would need to have a semi-major axis of 4 planetary Roche-radii for 100\% of the simulations to produce an observable moon. These values are comparable with the orbital properties of gas giant solar system satellites. We place similar constraints for eccentricities up to 0.1. We conclude that if  the semi-major axis and radius of the moon are known (eg. with exomoon transits), tidal dissipation can constrain the orbital eccentricity and interior properties of the satellite, such as the presence of melt and the thickness of the melt containing sub-layer.
}
{}


   \maketitle
%

\section{Introduction} \label{sec:intro}

Since all solar system giant planets host moon systems, it is plausible that exomoons orbit exoplanets as well. 
Even though detection techniques have shown tremendous successes in detecting over 5000 exoplanets, exomoon detection is more challenging because the deviation that is caused on the planet's signal by their presence is relatively small.
However, progress in observational astronomy has made the detection of exomoons a near-future possibility \citep{lazzoni2022,Teachey_kipping_2018, heller2017detecting, Heller_2016a}.

Several indirect and direct methods have been proposed to detect exomoons.
The effect that the moon has on the planet's signal is studied with indirect methods.
These include Transit Timing Variations \citep[TTV;][]{kipping2009,simon2007,Sartoretti_schneider_1999}, Transit Duration Variations \citep[TDV;][]{kipping2009}, and other photometric effects on the stellar signal \citep{heller2014}, centroid shifts \citep{Agol_2015} and Doppler monitoring of directly imaged exoplanets \citep{Ruffio_2023, Vanderburg_2018} caused by the gravitational interaction between the planet and its companion. 
Other methods for exomoon detection include dynamical sculpting of the circumplanetary disk \citep{Kenworthy_2015} and the detection of planet-satellite mutual eclipses \citep{cabrera_schneider_2007}.

Even though, there have been several tentative detections (see e.g., \cite{Kipping_2022, Oza_2019, Teachey_kipping_2018, Kenworthy_2015}), no exomoon detection has been confirmed yet, despite the numerous known exoplanets and the surveys searching for companions around them \citep{Kipping_2022, Kipping_2012}.
Apart from their small signal, this may also be due to the detection methods' biases to detect exoplanets closer to their host star, meaning that any potential satellite would be stripped off during the planet's migration (see e.g., \cite{Dobos_2021, Trani_2020}).

Direct imaging, which is sensitive to young exoplanets further out from their host star, could offer an observational window to far out systems.
Only recently, \cite{Benisty_2021} directly imaged the thermal emission from a circumplanetary disk at sub-millimeter wavelengths in the PDS 70 system, a strong prerequisite for exomoon formation. 
Direct imaging of THEMs is challenging, because except for the need to suppress the star's light, the final signal contains both the flux from the moon and the planet. However, Tidally Heated Exomoons \citep[THEMs;][]{Limbach_2013} can be brighter than their host planet's signal in some infrared (IR) bands, due to tidal interactions between them and the host planet.
A direct detection via thermal Spectral Energy Distribution (SEDs), (in contrary to other proposed methods, for instance, transits) does not rely on a specific satellite orbital phase nor require multiple observations.
This method offers us a detection window for exomoons around gas giant planets at distances of several tens of Astronomical Units (AU) from their parent star. It can also directly measure the flux emitted from the surface, which under the assumption of thermal equilibrium can be equal to the tidal heat flux produced in the interior.
Even though the mass of the satellite is not measurable via this method, the advantage of constraining the tidally generated flux can offer possibilities for characterization, because the latter depends on the interior structure and orbital properties of the moon. 

Tidal models over a broad range of complexity have been used to model solar system bodies and moons \citep{rovira2022, STEINKE2020113299,  Bierson_nimmo_,   Castillo-Rogez_2011,Hussmann_Spohn_2004,FISCHER_1990,  REYNOLDS1987125} and have already been applied to exoplanets \citep{Barr_2018,Shoji_Kurita_2014,Henning_2009}.
For the first time, \citet{Dobos_2015} modeled THEMs taking into account viscoelasticity, a behavior that describes the tidal response of the material making up the interior of planetary bodies. 
\citet{rovira_2021} used a coupled thermal-orbital evolution model and compared the effect that different rheological models have on THEM surface temperatures, while studying their longevity.
Finally, \citet{jager2021} took into account the presence of a hotspot on the surface of homogeneous THEMs without melt and modeled the increased radiation in thermal wavelengths to infer detectability.

In this work, we study the effect of tidal heating on exo-Ios' direct imaging observations. We define an exo-Io as an exomoon with the same density and interior structure compatible with Io's, with a thin, melt-containing asthenosphere beneath the lithosphere \citep{MOORE2001}. 
We use models developed to describe Io's tidal heat flux \citep{rovira_2021,Moore_2003}, and apply them to THEMs with the assumption of thermal equilibrium. We use $\epsilon$ Eridani b as a test case for a potential host planet. $\epsilon$ Eridani b is a close-by gas giant exoplanet with a distance at 3.2 parsec \citep{MacGregor_2015} and a minimum mass of 0.65 - 0.78 $M_J$ \citep{ 2021ApJS..255....8R,2021AJ....162..181L, Mawet2019}, making it a plausible candidate to host relatively large satellites \citep{Canup2006}. 
In Section~\ref{sec:model} we explain the tidal-thermal model and the interior structure used in this work, as well as the algorithm that solves for the surface temperature of moons in thermal equilibrium.
Section \ref{sec:observ} describes how we define an observable moon around $\epsilon$ Eridani b.
In Section \ref{sec:sectionresults} we present the orbital-interior configurations that would lead a THEM around $\epsilon$ Eridani b to be detectable with MIRI, under the assumptions of our model. We then extend this analysis with a sensitivity analysis over the various free parameters of our interior model, allowing us to quantify which interior constraints a detection may provide. Finally, in Section \ref{sec:discuss} we discuss possible ways to validate our analysis, factors that could affect the interpretation of our results, such as the presence of moon atmospheres and different moon sizes, as well as exomoon constraints that our method can yield in combination with other exomoon detection methods.

\section{Model}\label{sec:model}

\begin{figure}[!htb]
    \centering  
    \includegraphics[width=\columnwidth]{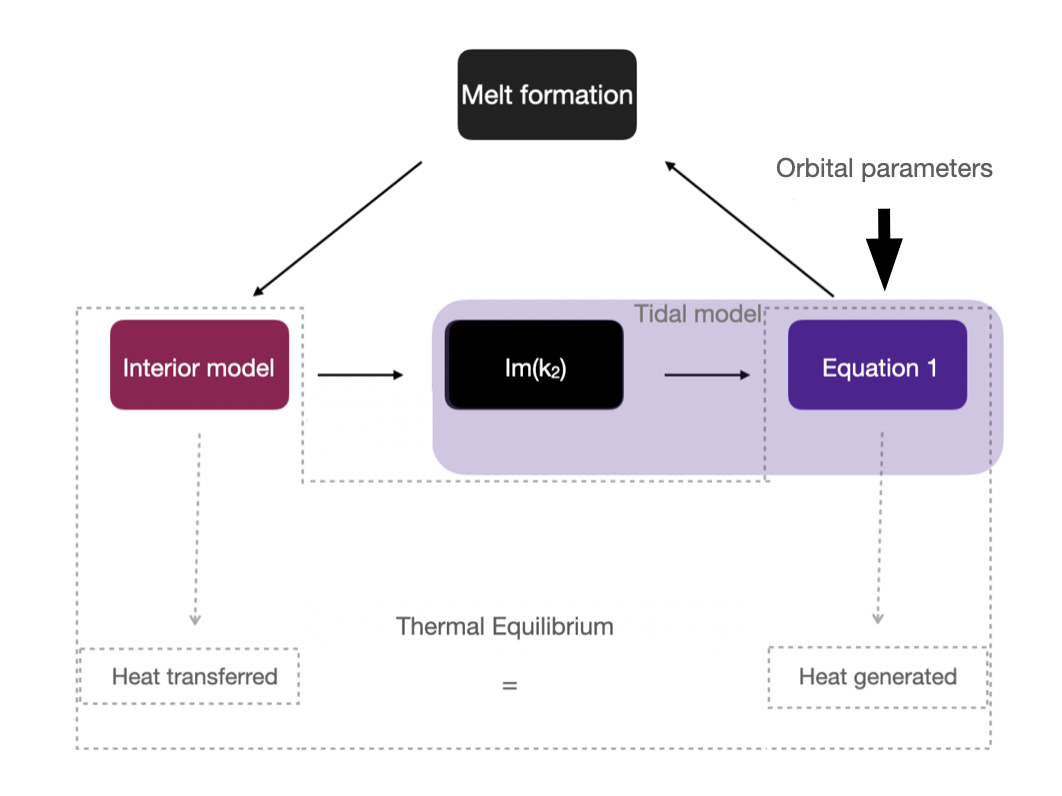}    
    \caption[Draft model]{Feedbacks between the heat transfer and the tidal models. The interior structure affects the \begin{math}Im(k_2)\end{math} value, on which the tidal dissipation depends on. High tidal dissipation can lead to the formation of melt, forming a new sub-layer (new interior structure) (Upper part).
    Thermodynamic equilibrium is reached when the moon's interior transfers the same amount of heat as the one that is generated via tidal interactions (Lower part).}    
    \label{fig:feedbacks}   
\end{figure}

To evaluate a THEM's detectability and to infer its surface heat flux, we use an interior model, which describes the moon's layers and heat transfer mechanisms, and a tidal model, which calculates the tidal dissipation given the interior structure and orbital parameters. A schematic overview of how these two models interact is given in Figure \ref{fig:feedbacks}.

We model THEMs using models initially developed to describe Io's interior, which have been shown to be compatible with Io heat flux observations \citep{spencer2000, veeder1994}. In our model, we assume a metallic core, solid silicate mantle and a partially molten sub-layer of melt beneath the stiff lithosphere. The moons are assumed to be spherically symmetric and the properties of each layer are uniform. We assume that tidal heat is generated in the deep mantle and asthenosphere (viscoelastic layers) and that the lithosphere behaves elastically. Our model implementation is adapted from \cite{rovira_2021}

In the following subsections we introduce both the interior and tidal models used.

\subsection{Interior model}\label{interiormodel}

\begin{figure}[!htb]
    \centering  
    \includegraphics[width=0.8\columnwidth]{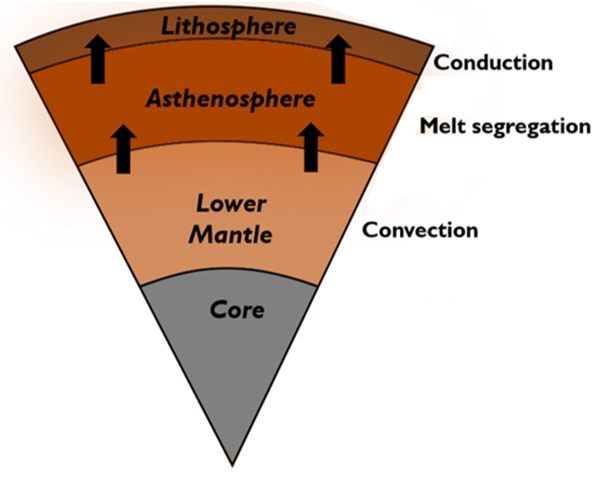}   
    \caption[thermal model]{Interior layers and the corresponding heat transfer mechanisms assumed in our model}    
    \label{fig:thermalmodel}   
\end{figure}

We assume different heat transfer mechanisms through the interior layers (see Figure~\ref{fig:thermalmodel}). The heat that is generated through tidal friction in the viscoelastic layers (mantle, asthenosphere) is ultimately emitted via radiation from the surface. We assume that the heat is transported via convection in the lower mantle, and conduction in the stiffer lithosphere, compatible with bodies in a stagnant lid regime (eg. \cite{REESE1999, SCHUBERT1979}).

For high tidal heating rates, the interior temperature rises and a partially molten sublayer, the asthenosphere, forms beneath the moon's lithosphere. We model the heat transport in the asthenosphere, as introduced by \cite{MOORE2001}, with melt segregation. This model successfully describes Io's current state. If no melt segregation is assumed, convection alone cannot explain the observed heat flux of Io, assuming thermal equilibrium \citep{Moore_2003}. As such, the Super-Ios discussed in our work are modeled with the same mechanism.

\begin{figure}[!htb]
    \centering  
    \includegraphics[width=\columnwidth]{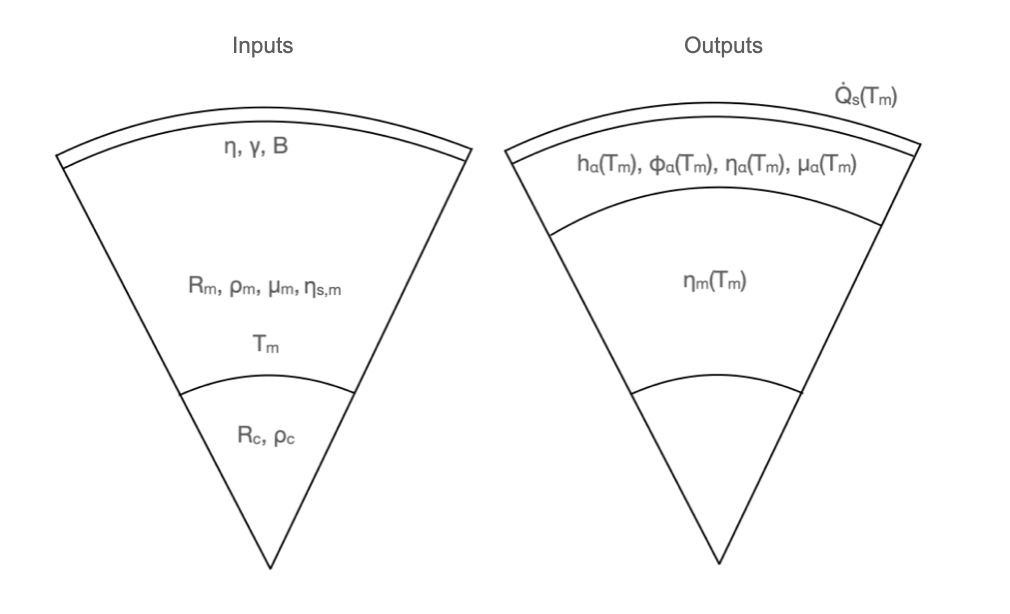}   
    \caption[parameters]{Input and output parameters used in the thermal model. Descriptions of these values can be found in Tables \ref{table:interior inputs}, \ref{table:interior parameters} and \ref{table:sensitivity parameters}}   
    \label{fig:parameter}   
\end{figure}

Overall, the parameters of the interior model are shown in Figure \ref{fig:parameter}.
We assume a core of radius $R_c$, equal to $ 0.52\times{R}$ \citep{anderson}, where R is the moon radius, with density $\rho_c$, a mantle with density $\rho_m$, shear modulus $\mu_m$, solidus viscosity $\eta_{s,m}$ and a temperature $T_m$.The dependence of the mantle viscosity ($\eta_m$) on the temperature is parameterized with the activation energy $E_a$ (Equation \ref{eq:viscositymantle}). The asthenosphere's viscosity ($\eta_{\alpha}$) also depends on the presence of melt through the $B$ parameter, as shown in Equation \ref{eq:viscosityastheno} \citep{MEI2002}.
Finally, the melt segregation is a function of the permeability exponent $n$ and the scale velocity $\gamma$. 
\begin{subequations}\label{eq:viscosity}
    \begin{align}
    \eta_m = \eta_{s,m}  exp \left(\frac{E_a}{R_gT_s}\left(\frac{T_s}{T_m} - 1 \right) \right) 
    \label{eq:viscositymantle}
    \end{align}
      
     \begin{align}
     \eta_{\alpha} = \eta_{s,m}  exp \left(\frac{E_a}{R_gT_s}\left(\frac{T_s}{T_m} - 1 \right) \right) exp\left(-B\phi_\alpha\right)
     \label{eq:viscosityastheno}
     \end{align}
\end{subequations} 
where $R_g$ is the ideal gas constant, $\phi_\alpha$ the melt fraction of the asthenosphere, and $T_s$ the average of the mantle solidus temperature.
The input values used can be found either in Table \ref{table:interior inputs} or in Table \ref{table:sensitivity parameters} for the parameters included in  our sensitivity analysis (see Section \ref{sec:sensitivity}) .

\begin{table}[htb]
\caption{Input interior parameters}
\centering
\begin{tabular}{c c c c c}
\hline\hline
Parameter & Symbol & Value & Units \\ [0.5ex] 
\hline
Core radius & $R_c$ & $0.52 \times R $\\
Mantle density & $\rho_m$ & $3542$ &  $kg m^{-3}$\\

Core density & $\rho_c$ & $5150$ & $kg m^{-3}$ \\
Activation energy & $E_a$ & $300$ & $kJ mol^{-1}$ \\
\hline
\end{tabular}

\label{table:interior inputs}
\end{table}

\begin{table}[htb]
\caption{Interior parameters - outputs }
\centering
\begin{tabular}{c c c c c}
\hline\hline
Parameter & Symbol \\ [0.5ex] 
\hline
Asthenosphere melt fraction& $\phi_\alpha$\\
Asthenosphere thickness & $h_\alpha$ \\
Asthenosphere viscosity & $\eta_\alpha$\\
Asthenosphere shear modulus &$\mu_\alpha$  \\
Mantle viscosity & $\eta_m$ \\
Mantle temperature & $T_m$\\
[1ex]
\hline
\end{tabular}
\label{table:interior parameters}
\end{table}

The interior model output parameters, which are obtained once thermal equilibrium is reached (also listed in Table \ref{table:interior parameters}) consist of the asthenosphere thickness ($h_{\alpha}$) and melt fraction that are needed to transfer $\Dot{Q_s}$ heat via melt segregation for a value of $T_m$. The material properties that make up the moon's interior are temperature dependent, thus all the output parameters, listed in Table \ref{table:interior parameters} and Figure \ref{fig:parameter} are a function of $T_m$. \cite{rovira_2021} explain in detail the interior properties dependency on the melt fraction and temperature, and they include a detailed description of the heat transfer model (including convection and conduction). We highlight, that when the mantle temperature crosses a threshold, which corresponds to the fraction of melt in the asthenosphere rising above 0.45 \citep{Moore_2003}, the rheology of the mantle is no longer adequately described using viscoelasticity, and is more accurately portrayed as a magma ocean, a regime which is not modeled in our work. We only discuss the relevance of a magma-ocean regime qualitatively in terms of detectability.

\subsection{Tidal Model}\label{tidalmodel}
The amount of tidal heat dissipated in the interior of a moon with zero obliquity is given (eg. \cite{Segatz_Spohn_1988, Makarov_2014}):

\begin{equation}
    \Dot{E} = -\frac{21}{2} \text{Im}(k_2)  \frac{\left(nR\right)^{5}}{G} 
    e^2 
 \label{eq:tidalflux}
\end{equation}   
where $n$ is the mean motion, $e$ the orbital eccentricity of the satellite, and G the universal gravitational constant. We are assuming synchronous rotating moons and perform simulations for eccentricity values up to 0.1. For larger eccentricities, high order terms ($\mathcal{O}(e^4)$) must be taken into account, since Equation \ref{eq:tidalflux} is accurate until second eccentricity order \citep{Renaud_2021}. We assume that the moon has zero obliquity, close to Io's \citep{Baland_2012}, and thus, we do not take it into account in the calculation of the total amount of tidal heat generated.

$\text{Im}(k_2)$ is the imaginary part of the $k_2$ Love number, which describes the  deformation of the moon's gravity field due to the tidal interactions with the planet. 
It depends on the interior structure, properties, rheology of the moon and it's orbital period \citep[] {Segatz_Spohn_1988}.
Efforts to obtain $\text{Im}(k_2)$ for Io \citep{Lainey2009} constrained it at 0.015 +/- 0.003 with astrometric observations.
 
Equation~\ref{eq:tidalflux} demonstrates how the tidally generated heat flux of a moon depends on the orbital properties of the satellite.
Self-luminous exomoons closer to the host planet and with higher orbital eccentricity would produce higher tidal heat fluxes, favoring their observability.

\subsubsection{Different approaches of $\text{Im}(k_2)$ calculation}

A simplistic approach to calculate the imaginary $k_2$ Love number assumes that $\text{Im}(k_2) = \frac{k_2}{Q} = constant$, where $Q$ (tidal quality factor) quantifies the fraction of the orbital energy that is dissipated per orbit due to friction.
This parameter is not well constrained, however rocky bodies are expected to have values from 10 to 500 \citep{GOLDREICH1966375}. 
Such an approach, does not take into account the feedback between the thermal state of a planet and its response to tidal forces.
In addition, it doesn't consider the tidal quality factor's dependency on the orbital period of the satellite \citep{Renaud_Henning_2018}.
It has also been found to underestimate tidal heat production (see eg. \cite{MEYER2007535, Dobos_2015}). 

Another approach, which has been widely used for solar system moons \citep{FISCHER_1990,Hussmann_Spohn_2004}, is to model tidal dissipation using the viscoelastic theory for self-gravitating bodies \citep{Peltier_1974, sabadini}, where a rheological law is needed to couple stress and strain. This method allows for the incorporation of the material properties' temperature dependence (eg. \cite{Karato}). The most simple viscoelastic model is the Maxwell model. The value of  $\text{Im}(k_2)$ is closely related to the Maxwell time ($\frac{\eta}{\mu}$). When the tidal period is close to this value, tidal dissipation reaches its maximum. For periods longer than what corresponds to the Maxwell time, the moon responds as a viscous fluid and for shorter tidal periods, as an elastic body. This approach has also been implemented for the study of exoplanet tidal responses \citep{Shoji_Kurita_2014,Henning_2009,Barr_2018}.

A viscoelastic approach to describe tidal heating in exomoons was used for the first time by \cite{Dobos_2015}.
They implemented the Maxwell viscoelastic model, which, however, does not accurately describe the laboratory behavior of olivine \citep{JACKSON_2010} and does not take into account the anelastic transient creep response over timescales shorter than the Maxwell time. The transient creep is described as a regime where the strain rate is a function of time.
On the contrary, the (more) advanced Andrade rheological model \citep{Andrade_1910} adopts this behavior and is more realistic.
The Andrade model 
 has been used in studies of solar system bodies \citep{Castillo-Rogez_2011,Bierson_nimmo_} and exoplanets \citep{Renaud_Henning_2018, walterova_2020}. 
 \cite{Renaud_Henning_2018} found significantly higher tidal heating when assuming Andrade rheology for Io-like moons,  thus the incorporation of it -- instead of the Maxwell one -- has implications on THEM detectability.

\subsubsection{Calculation of $\text{Im}(k_2)$}
We use the viscoelastic theory for self-gravitating bodies, incoorporating the Andade model to calculate $\text{Im}(k_2)$.
 The tidal response of the moon can be obtained by solving the equations of motion for the deformation of each layer in the Fourier domain, via the correspondence principle \citep{Peltier_1974}. The latter results in a set of differential equations that describe the deformation of each layer.

To solve the differential equations and obtain the gravitational potential at the surface, we use the propagator  matrix technique \citep{JARAORUE2011417, sabadini} and the Andrade rheological law. For more details on how we solve the propagator matrix technique see Appendix A of \cite{rovira_2021}. The viscoelastic response of the material in each layer depends on the shear modulus $\mu$ and viscosity $\eta$. 
The Fourier transformed shear modulus $\tilde{\mu}$ is related to the creep function $\tilde{J}$, which for the Andrade law is \citep{Efroimsky_2012}:

\begin{align}
    \tilde{\mu} &= \tilde{J}^{-1}
    \label{eq:creep}\\
    \tilde{J} &= \frac{1}{\mu} - \frac{i}{\eta n} + \frac{\mu^{\alpha-1}}{(i\zeta\eta n)^\alpha } \alpha!
        \label{eq:andrade}
\end{align}   

The transient creep response is modeled in the last term of equation \ref{eq:andrade} and is described by two parameters
$\zeta$ and $\alpha$. We assume that $\zeta = 1$ for the rest of our work. For assumptions on $\alpha$ see Section \ref{sec:sensitivity}.

\subsection{Thermal equilibrium}\label{sec:equilibrium}
The tidal model describes the heat that is generated through tidal interactions and the heat transfer model simulates how this energy gets transferred through the different layers of the interior. The two models are not independent, but interact with each other through feedbacks (Figure \ref{fig:feedbacks}, upper part); an $\text{Im}(k_2)$ value corresponds to a particular interior structure, but also affects the amount of tidal dissipation that is produced (Equation \ref{eq:tidalflux})). For high values of tidal dissipation, the interior temperature increases, and consequently the interior properties, layers and melt fraction change. This means that the interior structure is modified, affecting the $\text{Im}(k_2)$.

To this end, equilibrium is reached once the amount of heat emitted from its surface into space is equal with the one that is generated (Figure \ref{fig:feedbacks}, lower part).
This loop ends, when the moon reaches thermal equilibrium. \cite{Henning_2009} showed that significantly tidally active planetary bodies achieve thermal equilibrium in a few million years.  The overall thermal equilibrium equation that we solve for is:

\begin{equation}
    \Dot{Q_{tid}} - \Dot{Q_{s}} = 0, 
\end{equation}
where $\Dot{Q_{s}}$ is the surface heat flow, $\Dot{Q_{tid}}$ the internal heat production. $\Dot{Q_{s}}$ is computed via the equations for melt segregation presented in \cite{Moore_2003} and $\Dot{Q_{tid}}$ via Equation \ref{eq:tidalflux}.

In thermal equilibrium, the average surface temperature $T_{surf}$
 can be calculated by Stefan-Boltzmann's law:

\begin{equation}
 T_{surf}^4 = \frac{q_s + \frac{(1-A)S}{4}}{\sigma} \label{eq:tsurf}
\end{equation} 
where $\sigma$ is the Stefan-Boltzmann's constant, $q_s$
the equilibrium surface heat flux (where $q_s=\dot{Q}_{s}/(4\pi R^2$), A the moon's bond albedo (assumed equal to Io's) and S the stellar irradiation. For $\epsilon$ Eridani  this corresponds to $0.34 \times L_{Sun}$ \citep{Saumon_1996}, where $L_{Sun}$ is the solar luminosity. The second term of Equation \ref{eq:tsurf} describes the heat flux that is not reflected by the moon's surface and, thus, is taken into account in the thermal budget. 

 \begin{figure}[!htb]
    \centering  
    \includegraphics[width=0.9\columnwidth]{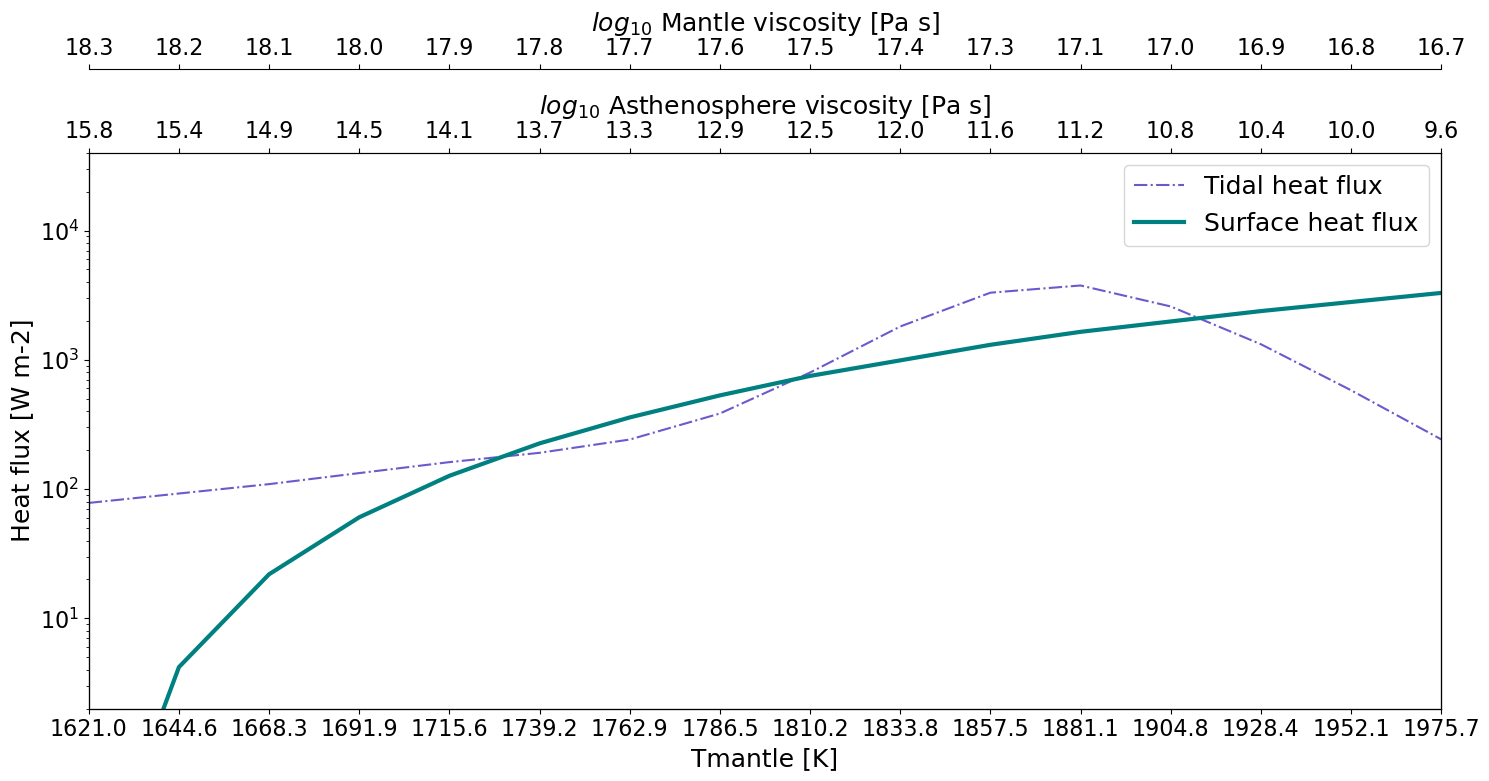} 
    \caption[Stable equilibria]{Generated tidal heat flux as a function the mantle temperature ($T_m$) and the log of the asthenosphere's ($\eta_{\alpha}$) and mantle viscosity ($\eta_{m}$). Stable  and non-stable equilibria points are shown as black and grey dots correspondingly. At a non-stable equilibrium point, a deviation in the mantle temperature drives the moon out of equilibrium, whereas a body at a stable thermal equilibrium state tends to restore its equilibrium. The tidal heat flux is calculated for a moon with semi-major axis $a = 6.59$ Roche radii, e = 0.02, $\alpha$ = 0.5, B = 30 and the rest of the sensitivity parameters set to their Nominal values (Table \ref{table:sensitivity parameters})}   
    \label{fig:stable}  
\end{figure}

Thermal equilibrium points are defined as the intersections of the tidal and the transferred heat (i.e. the heat generated and the heat advected to the surface of the moon). These can be stable or non-stable \citep{Moore_2003}, as shown in Figure \ref{fig:stable}. At a stable equilibrium point, a temperature increase would lead to a higher heat advection rate compared to the heat generation one. The moon, thus, tends to restore its equilibrium. The reversed scenario is described by a $\frac{d\Dot{Q_{s}}}{dT_m} < \frac{d\Dot{Q_{tid}}}{dT_m}$. In this case, a slight increase in the interior mantle temperature would drive the moon out of equilibrium, since the heat production would exceed the advection capabilities of the moon, warming up the interior (Figure \ref{fig:stable}). High temperatures coming from formation would potentially make the moon approach Figure \ref{fig:stable} from the right side of the graph. This is a favorable condition for the moon to firstly reach the higher stable equilibrium states, such as the warmer stable equilibrium point.

In this work, we present results for the stable equilibrium points that correspond to higher temperatures. Scenarios where the moon would reach the lower stable equilibrium point exist (e.g., going out and in an MMR -- see \cite{fuller2016}), however studying these cases is beyond the scope of this work, which is mainly to assess whether and for what interior and orbital properties putative THEMs could reach temperatures that allow observations and under what assumptions we can place constraints on their properties.

\section{Observability}\label{sec:observ}

\begin{figure}[!htb]
    \centering  
    \includegraphics[width=0.9\columnwidth]{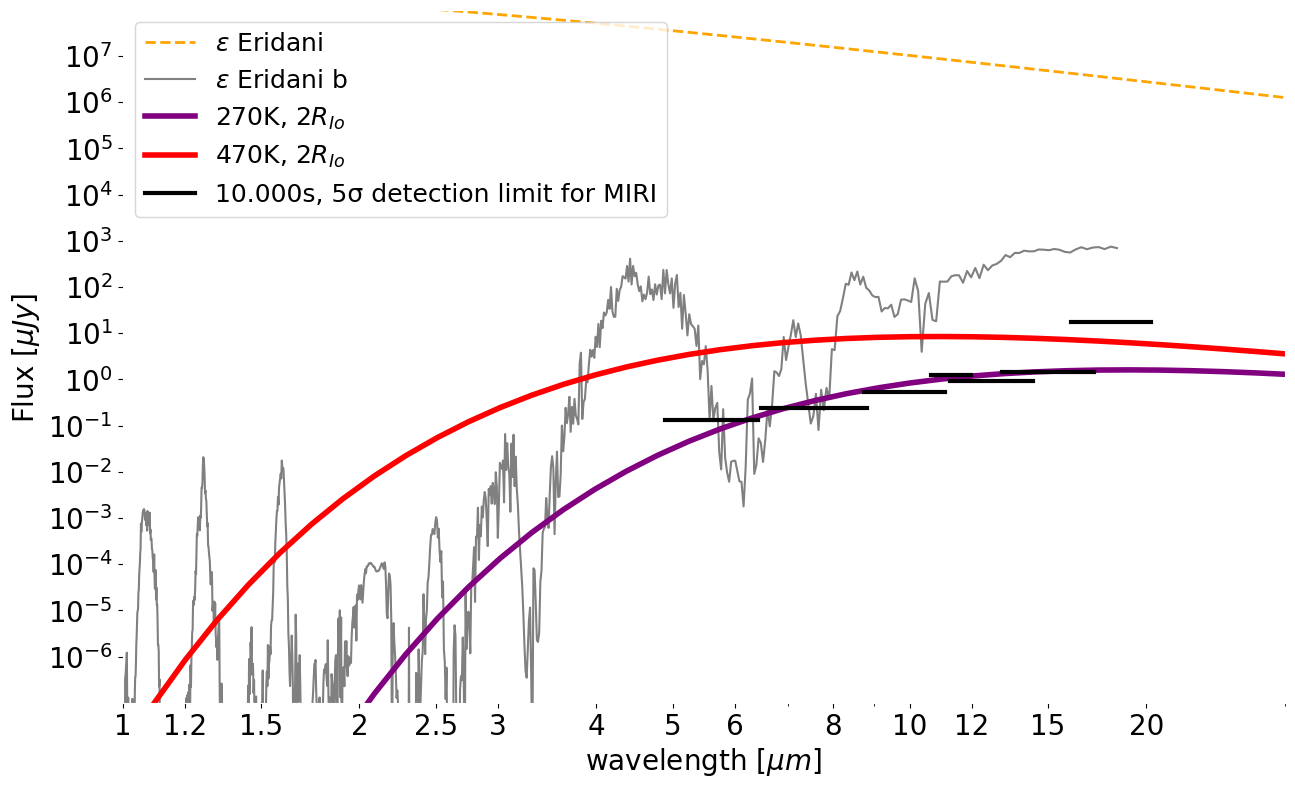}   
    \caption[Exomoons beta pic]{Fluxes of hypothetical $2 R_{Io}$ moons around $\epsilon$ Eridani b.
    The black lines are the 10 $\sigma$, 10,000 second detection limits for MIRI \citep{Glasse_2015}. The planet model \citep{Spiegel_2012} is shown in grey. The red and purple lines correspond to the black bodies of a 270 and 470 K $2 \times R_{Io}$ exomoon and the yellow dashed line to the black body of the star.}
    \label{fig:exomoonseeri}   
\end{figure}

In this section we define what we consider to be an observable moon around $\epsilon$ Eridani b.
Figure \ref{fig:exomoonseeri} shows the expected star, planet and moon fluxes in this system.
$\epsilon$ Eridani b is an exoplanet that has been indirectly detected through radial velocity with a mass of $\approx 0.65 M_{J}$ \citep{2021ApJS..255....8R,2021AJ....162..181L}.
The planet's spectrum is approximated via the models for young gas giants of \citep{Spiegel_2012}, assuming no clouds, 500 Myr age, 1 $M_J$ and solar metalicity ($\epsilon$ Eridani b: -0.04 dex)\citep{2021ApJS..255....8R}.
The star's SED is shown as a Blackbody of temperature 5084 K \citep{kovtyukh}.
In some wavelength regions the flux of the THEM surpasses that of the planet - for a 270K exomoon, these are at 3.6 and 6 microns, which widen significantly at higher exomoon temperatures.

MIRI is equipped with 4 coronagraphs; 1 Lyot and 3 four-quadrant phase masks \citep[4QPMs; ][]{Boccaletti2022}.
Of these, the 4QPMs have the smallest achievable inner working angles (IWAs), between 0.33 and 0.49 arcseconds \citep{Boccaletti_2015}.
 For a semi-major axis of 3.5 AU \citep{2021AJ....162..181L}, the planet has a separation of $\approx$ 1.09 arcseconds from its host star - outside of the inner working angle of the 4QPM coronagraphs.
\cite{Boccaletti2022} measured the on-sky performance of MIRI's coronagraphs and concluded that post-processing techniques can bring the final contrast down to the background and detector limited noise floor at separations larger than 1 arcsecond. Future instruments, like METIS \citep{brandl2021}, are expected to reach even smaller contrast ratios.

We therefore define a ``detectable exomoon'' as the coldest exomoon that reaches the $10\sigma$ point source detection limit for MIRI in a  10,000 second integration at $\epsilon$ Eridani's distance, at 3.2 parsecs \citep{MacGregor_2015}.
The average moon surface temperature in our model is calculated through Equation \ref{eq:tsurf} and represents the theoretical total thermal output of the moon, assuming no localized variations due to volcanic activity or hotspots on the surface.
Figure \ref{fig:exomoonseeri} shows that this limit is reached for an average surface temperature of 270 K, for a $2 R_{Io}$ exomoon orbiting $\epsilon$ Eridani b.
Moons of the same size with a higher effective temperature are also considered ``detectable''.

The moon's thermal flux could be constrained from observed excess of heat flux at different wavelengths, where the planet is not expected to be bright, for example at water or methane absorption bands. It is beyond the scope of this work to disentangle the moon's signal from the planet's one in specific observational bands. It is rather to assess the detectabililty and orbital-interior configurations that can lead to observations. The use of different planetary atmosphere models could also potentially affect the combined planet-moon flux measured. We ignore any effect the latter would have in disentangling the putative detected flux in it's planet and moon components.

\section{Results} \label{sec:sectionresults}

In this section we aim at exploring how interior and orbital parameters could be constrained if the thermal heat flux of a THEM was known. We first explore how this could be known if our model was perfect and all the free parameters were well determined. Thus, the first subsection refers to results obtained assuming the nominal values of Table \ref{table:sensitivity parameters}. We then take into account modeling parameters uncertainties to explore how constraints can be placed, as shown in Section \ref{sec:sensitivity}. 

\subsection{Interior structure dependence on orbital properties}\label{sec:results}

\begin{figure}[!htb]
    \centering  
    \includegraphics[width=\columnwidth]{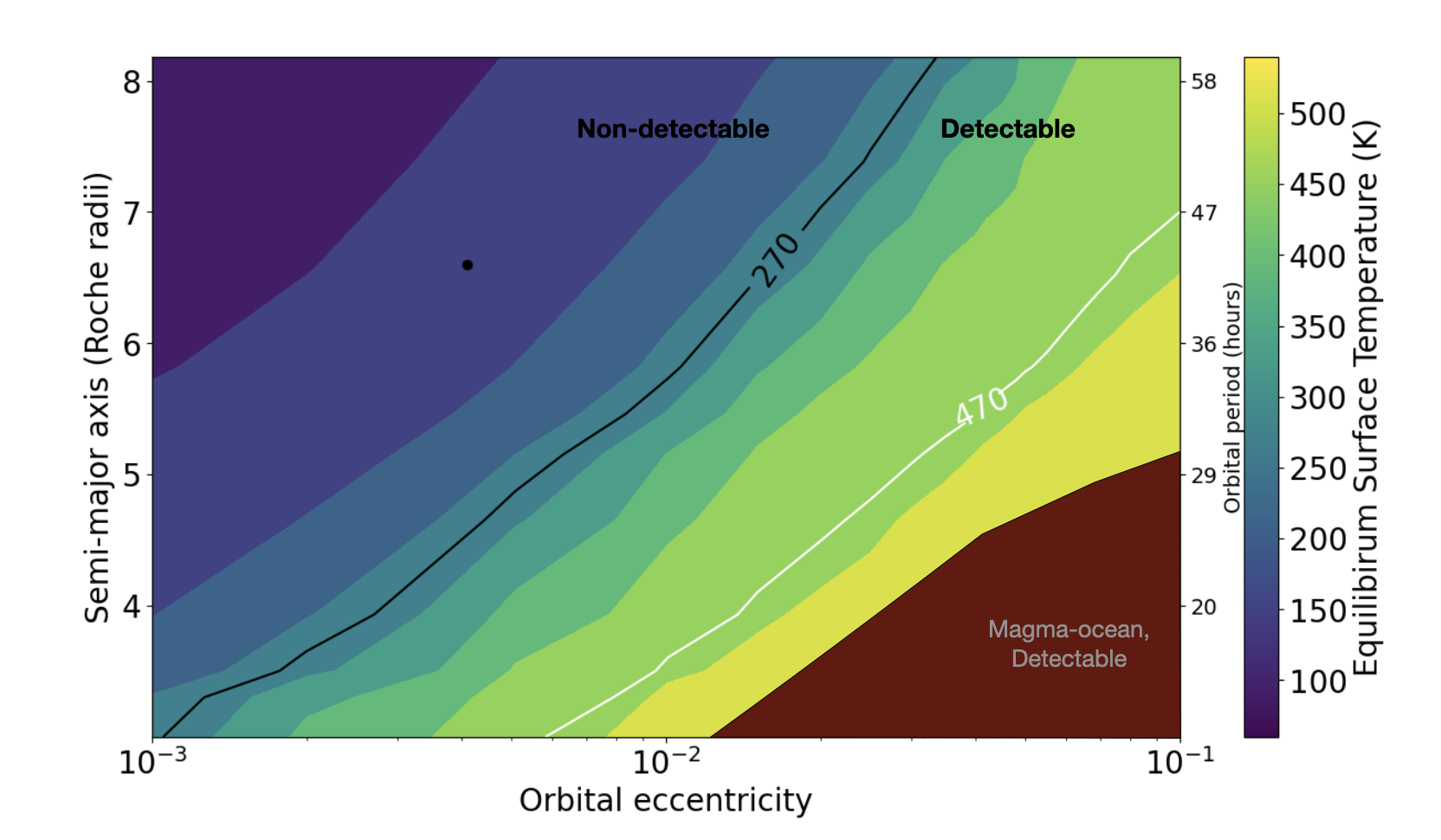} 
    \caption[Sketch3]{Surface equilibrium temperature of a 2R$_{Io}$ THEM as a function of its orbital eccentricity ($e$) and semi-major axis ($a$), Black dot: Io's current orbital parameters.
    The nominal values of interior parameters used are presented in Table \ref{table:sensitivity parameters}. The 270K isotherm divides the parameter space in Detecable and Non-detectable exomoons.} 
    \label{fig:Results:equilibriumtemp}
\end{figure}

Figure \ref{fig:Results:equilibriumtemp} shows the resulting surface temperatures using the models from Section \ref{sec:model}, for a $2R_{Io}$ exo-Io in a stable thermal equilibrium state as a function of the moon's semi-major axis $a$ and orbital eccentricity $e$.
For moons with the same density as Io this corresponds to $8 M_{Io}$.
Satellite formation theories indicate an upper limit of $M_{moon}/M_{planet} \approx 10^{-4}$ \citep{Canup2006}. 
According to this limit, exomoons of such mass could be found around 1.1 $M_J$ planets or higher.
This is compatible with the $\epsilon$ Eri b minimum mass between 0.65 and 1.55 $M_J$ \citep{2021ApJS..255....8R,hatzes}.
We focus our analysis on exomoons of the same size, because the latter needs to be constrained in order to derive any interior and orbital conclusions with our method.
The way to constrain the moon's radius, as well as the implications it will have in our results' interpretation are further discussed in Sections \ref{section: semimajoraxisconstr} and \ref{sec:largermoons}.

Tidal interactions between the host exoplanet and such exomoons could heat their surface temperature up to the order of $\sim 600$K for the shown ranges of orbital parameters (Figure \ref{fig:Results:equilibriumtemp}. Moons experiencing such an amount of tidal heating would be in a magma-ocean regime (see below).
The values of $a$ and $e$ that would make a Super-Io reach MIRI’s detection limit (Section \ref{sec:observ}) are demonstrated with the 270 K isotherm, which divides Figure \ref{fig:Results:equilibriumtemp} in two areas: ``Detectable'' and ``Non-detectable'' moons.
For example, a THEM orbiting epsilon Eridani b at 5.5 Roche radii would need to maintain an eccentricity higher than 0.009 to be detectable with MIRI.
This value is of comparable magnitude as solar system moons in MMR. %

Figure \ref{fig:Results:equilibriumtemp} can, thus, be used to predict surface equilibrium temperatures of THEMs under the assumptions of our model (see Tables \ref{table:interior inputs}, \ref{table:sensitivity parameters}).
The dark red area defines the area of the parameter space for which the melt fraction of the mantle exceeds 0.45 \citep{Moore_2003}. When this occurs, the asthenosphere behavior resembles a magma ocean. Tidal dissipation in such a regime is largely unstudied and depends on poorly constrained parameters.
Super-Ios with orbital characteristics that fall within this area produce higher tidal heat rates than what their interior can advect, either with convection or heat piping.
As a result, they are in a ``magma ocean'' state, where they reach the limits of our models, since liquid tides \cite{TYLER2011770} or poroviscoelasticity \cite{rovira2022} would need to be considered.
Nevertheless, such moons would be detectable (under our assumptions), since their surface would reach higher temperatures than those with lower asthenosphere melt. Because all moons with asthenosphere properties on the boundary of a magma ocean are detectable, moons with a magma ocean would be as well. Such moons could reach warmer thermal equilibrium states with a more efficient heat transport mechanism (convection in low viscous layer, \citep{solomatov}. They could also be still approaching thermal equilibrium at the moment of detection, radiating excess heat (eg. from formation).

In Figure \ref{fig:finalfig}a and b, the surface temperature and $\text{Im}(k_2)$ respectively of a Super-Io orbiting its host star at 4.65 Roche radii are plotted as a function of its orbital eccentricity in dark blue.
An assumption of a constant $\text{Im}(k_2)$ would lead to significantly different results compared to the correspondence principle technique.
Calculating $\text{Im}(k_2)$ via the correspondence principle technique as is done here, can affect our interpretation of whether an exomoon is detectable or not. For example, for the studied moon and a semi major axis of 4.65 Roche radii the equilibrium $\text{Im}(k_2)$ can change by a factor of 10 depending on the eccentricity (See Figure \ref{fig:finalfig}b dark blue line). From Equation \ref{eq:tsurf}) this difference would correspond to a calculated temperature that is off by 500 K, a temperature difference that as seen in Section \ref{sec:observ} can define whether a moon is observable.

\begin{figure}[!htb]
    \centering  
    \includegraphics[width=\columnwidth]{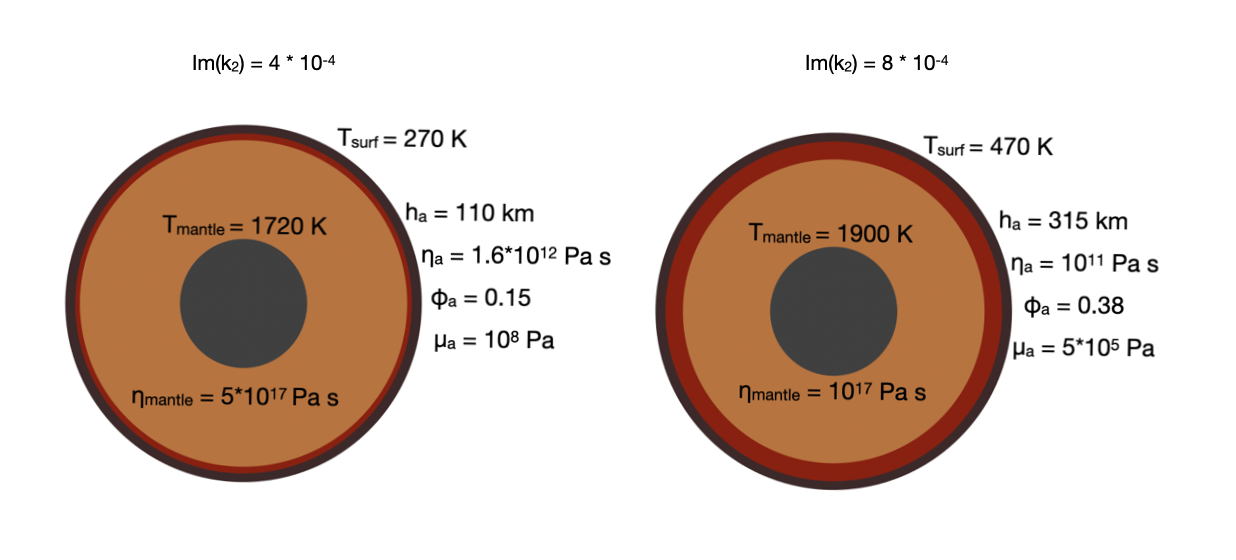}   
    \caption[example]{Two different equilibrium interior structures for $a = 4.65$ Roche radii.
    The one on the left ($e = 0.004$) has a thinner asthenosphere compared to the right one ($e = 0.02$) resulting to a smaller $\text{Im(}k_2)$.
    The parameter symbols are explained in Table \ref{table:interior parameters}.}
  
    \label{fig:2differentinteriors}   
\end{figure}

The value of $\text{Im}(k_2)$ provides a direct link between an exomoon observation with its interior structure, and could be used to draw conclusions on the interior properties shown in Table~\ref{table:interior parameters}. Figure \ref{fig:2differentinteriors} shows two different equilibrium interior structures that correspond to eccentricities of 0.004 and 0.02 and a semi-major axis of 4.65 Roche radii. The two different eccentricities have equilibrium surface temperatures of 270 and 470K respectively. The $\text{Im}(k_2)$ of the 270K exo-Io is approximately twice as high, for a semi-major axis of 4.65 Roche radii. 

Given that moons would reach stable thermal equilibrium at different tidal heat rates for each pair of $a$ and $e$, similar figures as Figure~\ref{fig:Results:equilibriumtemp} can be produced for every interior property in Table \ref{table:interior parameters}.
Figure \ref{fig:melt, vs e,w} shows how the melt fraction of the asthenosphere changes with varying $a$ and $e$.
\cite{jager2021} modeled THEMs with one hotspot, assuming non-homogeneous heat flux distributed amongst their surface.
Such hotspots enhance detectability in thermal wavelengths.
While our model does not take this into account, we highlight that a THEM with various hotspots could place constraints on its orbital period, from the variability of the signal the in direct imaging observations.

 \begin{figure}[!htb]
    \centering  
    \includegraphics[width=\columnwidth]{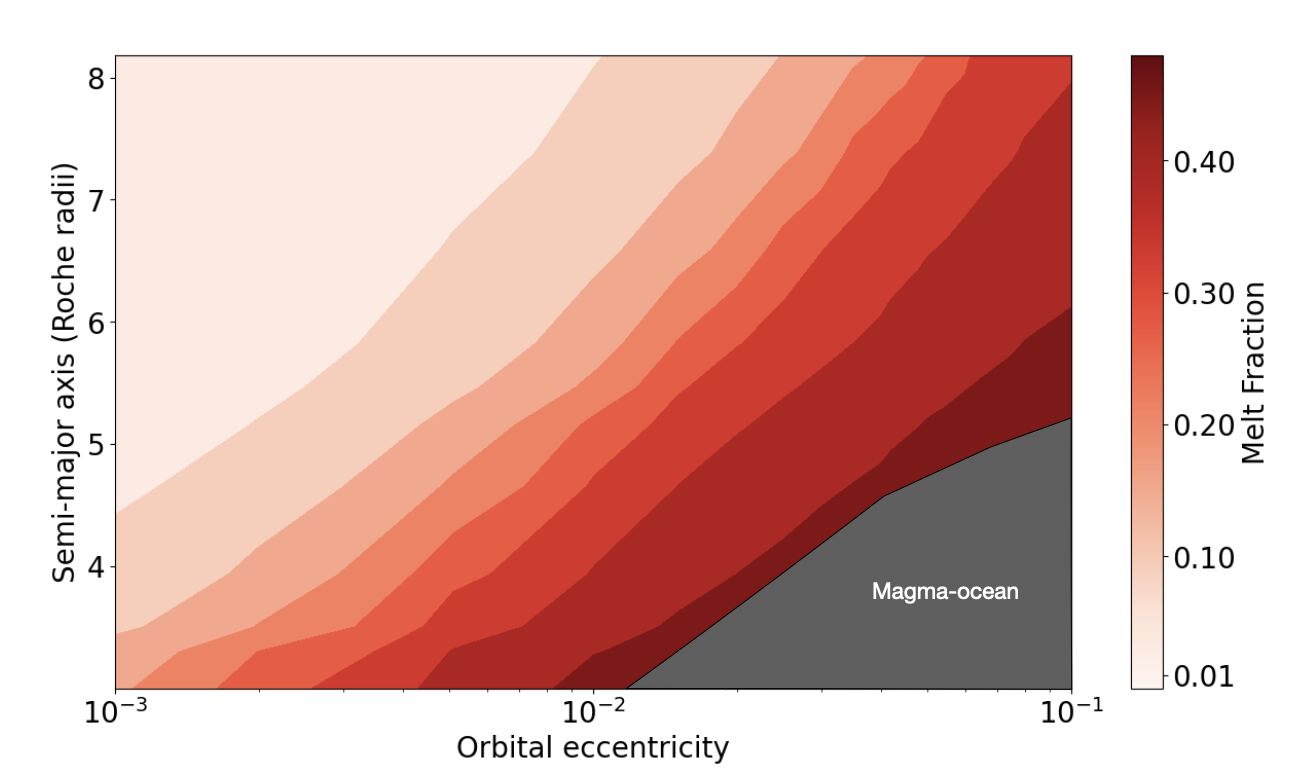}   
    \caption[example]{Asthenosphere melt fraction ($\phi_{\alpha}$) of stable thermal equilibrium states of exo-Ios as a function of their orbital eccentricity and semi-major axis. }  
    \label{fig:melt, vs e,w}   
\end{figure}

If the surface temperature and semi-major axis are known, the moon's eccentricity (Figure \ref{fig:Results:equilibriumtemp}) and interior properties (Figure \ref{fig:melt, vs e,w}) can be inferred, because a surface temperature value corresponds to one stable equilibrium $Im(k_{2})$ value for a given eccentricity.
In other words, the surface temperature of a putative exomoon with a given semi-major axis and interior structure would be a function of the orbital eccentricity. The dark blue lines in Figure \ref{fig:finalfig} show how, for a semi-major axis of 4.65 Roche radii, an inferred surface temperature (Figure \ref{fig:finalfig}a) and the interior properties (Figure \ref{fig:finalfig}b,c,d,e,f) change with the orbital eccentricity.
 Since some interior properties exhibit uncertainties, we performed a sensitivity analysis to asses whether and to what extend constraints on the orbital eccentricity and interior can be placed with a putative heat flux observation.

\subsection{Orbital and interior constraints from observations}\label{sec:sensitivity}

To summarize Section~\ref{sec:results}, our coupled tidal-thermal analysis can relate the surface heat flux of putative exo-Ios with the corresponding equilibrium $\text{Im}(k_2)$ values for a given moon semi-major axis.
This also means that constraints on the interior equilibrium states could possibly be deduced.
Given a semi-major axis, a single solution for the orbital eccentricity exists assuming the nominal values of Table~\ref{table:sensitivity parameters}.
However, the values of the parameters in Table \ref{table:sensitivity parameters} are not known exactly, and thus, their uncertainty needs to be incorporated in the conclusions that can be drawn with our model.

In this section we present the results of this analysis, varying the parameters presented in Table~\ref{table:sensitivity parameters} for the shown number of points and ranges. We obtain surface equilibrium  temperature as well as the corresponding interior parameters in Table \ref{table:interior parameters} with the goal to assess: 
\begin{itemize}
    \item how a surface temperature alone can constrain the moon semi-major axis and eccentricity 
    \item the fraction of simulations for a given eccentricity and semi-major axis that lead to an observable exo-Io 
    \item the constraints in 1) the interior properties and 2) the orbital eccentricity if the surface temperature and the semi-major axis are known
\end{itemize}
For all three purposes we vary the mantle shear modulus from $2\times 10^9$(\cite{STEINKE2020113299}) to $6.5\times 10^{10} Pa$, which corresponds to the end case of a mantle shear modulus equal to the lithospheric one assuming no melt \citep{Segatz_Spohn_1988,Peale1979}.
The Andrade parameter $\alpha$ can take values from 0.1 - 0.5 for olivine \citep{gribb1998}, while the melt fraction coefficient from 10 - 40 \citep{Henning_2009}. Finally, we vary the parameters that controls the magma advection between typical values ($\eta: 2-3$ \citep{Moore_2003}, $\gamma: 10^{-6} - 10^{-5}$ \cite{Katz}). Because the mantle solidus viscosity is highly uncertain, we vary it by a factor of 10 from the commonly used value of $10^{16} Pa*s$ \citep{ Renaud_Henning_2018, Moore_2003, FISCHER_1990}. We note that higher values of $\eta_{s,m}$ would need to be considered for internal structures with asthenosphere melt fraction larger than the treshold melt fraction defined in Section \ref{sec:model} \citep{KERVAZO2022}.
Each set of the sensitivity parameters leads to different stable thermal equilibrium states, thus surface temperatures. 
The sensitivity parameters are varied altogether along the discussed ranges. Each pair of $a$, $e$ now can now result in 720 different equilibrium states.

\begin{figure}[!htb]
    \centering 
    \includegraphics[width=0.8\columnwidth]{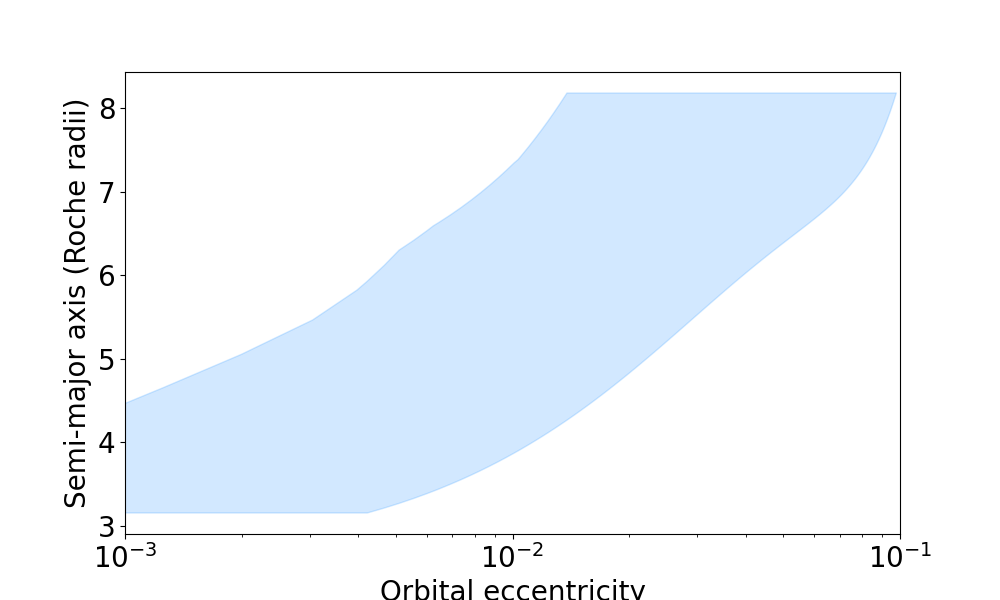}   
    \caption[Sketch2]{Range of exo-Io orbital properties that are compatible with a surface temperature of 270K and a radius of $2 R_{Io}$ as a result of the sensitivity analysis.}  
    \label{fig:range270}   
\end{figure}

The influence of the sensitivity analysis on the final equilibrium temperature is demonstrated in Figure~\ref{fig:range270}, where the $a,e$ values that can lead to 270 K (observable with MIRI, Section~\ref{sec:observ}) is shown.
This already puts constraints on the orbital properties that a particular heat flux observation would correspond to.
For example, an exo-Io with an eccentricity of 0.01 and a surface temperature observation of 270 K, could lie between $\approx$4-7.2 Roche radii away from the host planet in thermal equilibrium.

\begin{table*}[t]
\caption{Parameters and their ranges used for the sensitivity analysis.}
\centering
\begin{tabular}{c c c c c}
\hline\hline
Parameter & Description & Nominal value (Section 5) & Range (Section 6) & Number of points \\ [0.5ex] 
\hline
$\alpha$&Andrade parameter& 0.3 & 0.1 - 0.5 & 5 \\
$B$& Melt fraction coefficient & 20 & 10 - 40 & 4 \\
$\mu_{mantle}$& Deep mantle shear modulus&  $6.5 \times 10^{10}$ Pa & $2 \times 10^9 - 6.5 \times 10^{10}$ Pa & 3 \\
$\gamma$ & Melt scale velocity & $10^{-5}$  &  $ 10^{-6} - 10^{-5}$ & 2 \\
$\eta$ & Permeability exponent & 2  & 2-3 & 2\\ 
$\eta_{s,m}$ & Mantle solidus viscosity & $10^{16}$ Pa s & $10^{15}$-$10^{17}$ Pa s &  3\\
\hline
\end{tabular}
\label{table:sensitivity parameters}
\end{table*}

\begin{figure}[!htb]
    \centering 
    \includegraphics[width=1\columnwidth]{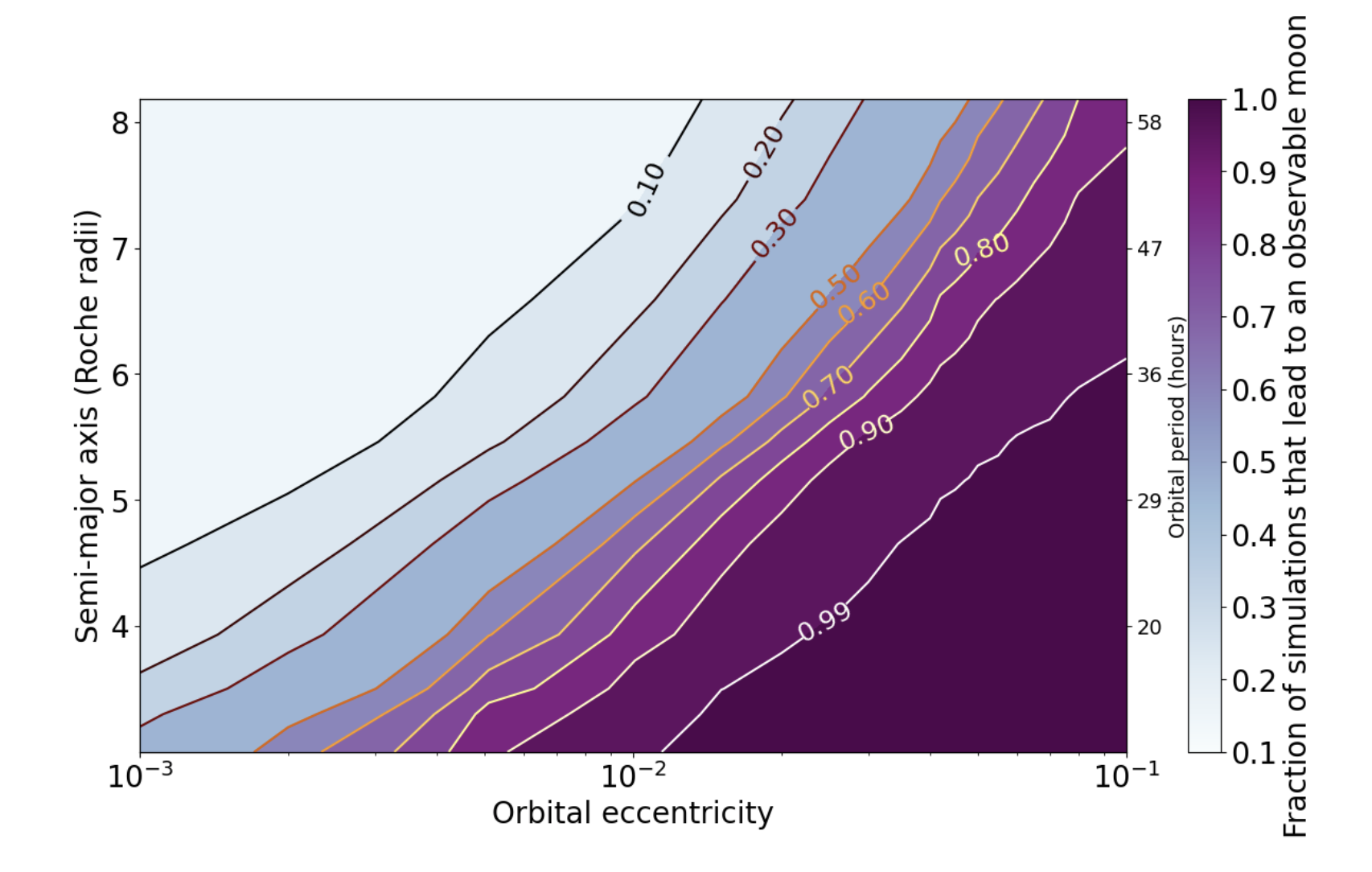}  
    \caption[Sketch2]{Fraction of simulations that lead to an observable moon (warmer than 270K (Section \ref{sec:observ})) as a function of its orbital semi-major axis and eccentricity of the moon.}    
    \label{fig:sensitivity_result}   
\end{figure}

To assess the fraction of simulations that result in an observable moon,we calculate the equilibrium surface temperature for the sensitivity parameters' ranges shown in Table \ref{table:sensitivity parameters}.
Each set of these parameters leads to a different equilibrium state for the same pair of $a,e$.
If the resulting surface temperature is higher than $270K$, we consider the moon observable (Section~\ref{sec:observ}).
We then calculate the fraction of the 720 simulated exo-Ios that are observable for each pair of semi-major axis and eccentricity and present the results in Figure \ref{fig:sensitivity_result}. 
For an eccentricity of 0.02, an exo-Io around $\epsilon$ Eridani b would need to have a semi-major axis of $\approx$4 Roche-radii for 100\% of the simulations to produce an observable moon.
While this eccentricity is about 5 times higher than the one of Io's and the semi-major axis is around 1.6 times smaller than Io's semi-major axis. 

Incorporating the uncertainties of the sensitivity parameters places more realistic constraints on the orbital eccentricity and interior properties for a theoretical $T_{surf}$ and $a$ detection.
Figure~\ref{fig:finalfig} shows the possible solutions  for the interior properties of an exo-Io orbiting $\epsilon$ Eridani b at 4.65 Roche radii as a function of the orbital eccentricity (light blue). 
The dark blue lines correspond to the nominal values analysis of Section~\ref{sec:results}.
For an observation of $270K$  (black dotted lines) only exo-Ios with $e$ smaller than 0.022 would be in thermal equilibrium.
This constraint is looser for higher temperatures.
For $470 K$ (red dotted lines) the corresponding range would be 0.01 - 0.1.

On the other hand, the constraints on the interior properties become stricter for higher temperatures. For example, only bodies with melt fractions larger than 0.1 can explain an observation of 470K for the given semi-major axis.  

As the eccentricity increases the equilibrium surface temperature increases, since the moon experiences more tidal heating.
As we approach larger eccentricities the range of possible surface temperatures (and the other interior parameters) tightens.
This is because the equilibrium $\text{Im}(k_2)$ 
experiences bigger variations close to the Maxwell time (see Figure \ref{fig:finalfig}b) for the considered range of parameters.
The peak of the curve, which corresponds to the Maxwell time, can be seen for lower eccentricities for the shown semi-major axis.
The equilibrium structure with viscosity and shear modulus tuned close to the Maxwell time for $a = 4.65$ Roche radii is formed for an eccentricity of 0.015 for the nominal values analysis.

\section{Discussion}\label{sec:discuss}

The direct detection of THEMs around exoplanets relies on disentangling the different fluxes (star, planet, moon) in a system at a given separation. The flux of the moon needs to be comparable (if not higher than) the planet's flux at a selected wavelength band, taking into account contrast limitations with the star. In this work we focus on modeling the moon's expected thermal flux, which depends on the amount of tidal heating that is produced in the moon interior, which is in turn a function of the interior structure and orbit of the moon. 

THEMs' expected tidal heat has been modeled following a more simplistic approach regarding tidal interactions and interior modeling by \cite{Limbach_2013}. For the first time \cite{Dobos_2015} took into account the viscoelastic behavior that describes the tidal response of the material making up the interior of planetary bodies. In their work \cite{Dobos_2015} included the Maxwell viscoelastic approach, commonly used to model the response of solar system planetary bodies \citep{SHOJI201310} and in exoplanet science \citep{Barr_2018}. They assumed a two layered body and convection as the main heat transfer mechanism in the mantle. In our analysis, we used a coupled thermal-tidal model combined with viscoelastic Andrade rheology and we assumed magma advection as the main heat transfer mechanism in a partially molten sub-layer. The effect of changing the classical Maxwell approach to the Andrade one has been studied for planetary bodies \citep{ Renaud_Henning_2018, Walterova2017}. In general, Andrade rheology produces higher tidal heating rates, due to the transient creep mechanism.  This is aligned with our results when compared with \cite{Dobos_2015}, who for a $2 R_{Io}$ moon at Io's orbital period and an eccentricity of 0.1 find 273 K. We find an equilibrium temperature above 400K for the same orbital properties. 

Tides provide a direct link of the interior with the orbital properties of a planetary body. Observations that can constrain the amount of tidal heat in the interior (e.g., thermal flux) combined with accurate modeling of the interior and potentially the orbit can help to infer system properties of exo-planet/moons. Direct observations of THEMs are (the only) direct method of probing the tidally heated surface. Our analysis shows the effect of tidal heating on the surface temperature of THEMs. We find regimes of orbital properties where the moon has to be in a magma ocean state to be able to dissipate the tidally generated heat. As seen in Figure \ref{fig:finalfig}, we also find that we can place constraints on interior properties and eccentricity depending on the (observed) temperature given a semi-major axis and a moon radius. Several ways of exomoon detection have been proposed (see Section \ref{sec:intro}), from which the exomoon mass  \citep{Kenworthy_2015,Sartoretti_schneider_1999,Vanderburg_2018,Teachey_kipping_2018}, radius and semi-major axis \citep{Teachey_kipping_2018} and atmospheric properties \citep{Agol_2015,Oza_2019} could be constrained.
Our work provides a way to infer system properties, like the orbital eccentricity and compatible interior structures.  
Finally a THEM detection would point out to the likely existence of at least a second moon in resonance, since tides would circularize an isolated moon's orbit.

In the following sections we explore some observational and modeling challenges. These include discussions on longevity, semi-major axis and radius constraints and more massive moons that could potentially host an atmosphere. Finally, we explore how our results could be potentially validated via other observational methods.

\subsection{Longevity and Mean Motion Resonance (MMR)}
$\epsilon$ Eridani is a relatively young system \citep[400-800 Myr; ][]{Mamajek_2008,janson2015}.
Such systems have potentially a disadvantage for the direct detection of THEMs; any planet in the system is expected to be warm from formation.
Since the planet-moon contrast also plays a significant role in THEM observability, this could  hinder (possibly) a detection. 

On the other hand, the system's young age could offer an observational advantage compared to older systems.
In an isolated moon-planet system tides quickly circularize the moon's orbit and close moons can migrate outward quite fast \citep{rovira_2021}, which is limiting the observability window. A second moon in resonance can boost the eccentricity of the first moon and maintain the surface temperature high enough for detection via tidal heating.
In both Jupiter's and Saturn's satellite systems, there are currently MMRs that likely originated with the outward migration of inner satellites, capturing the outer ones in MMRs \citep{YODER19811,Dermott1988}. 
Even though the MMRs in the solar system indicate that they are likely prevalent in older systems as well, it is reasonable to assume that MMRs are common in relatively young systems, since moons can be captured in MMRs during their formation \citep{peale2002,Ogihara_2012}.

The lifetime of an MMR depends on the mass ratio of the moons, among other parameters, and a short lived resonance could last $<$200\,Myr \citep{Tokadjian2022}. \cite{rovira_2021} calculated that an exo-Io with $2R_{Io} $ could have a temperature higher than 400K for 10 Myrs (during the early stages of an MMR). This would make such a moon detectable for 10 Myrs. Thus, if one considers the expected timescale of MMRs, young systems could potentially serve as good targets for THEM observations. 

\subsection{Combination with other exomoon detection methods: constraining the moon radius and semi-major axis} \label{section: semimajoraxisconstr}

\begin{figure*}
    \centering 
    \includegraphics[width=0.7\columnwidth]{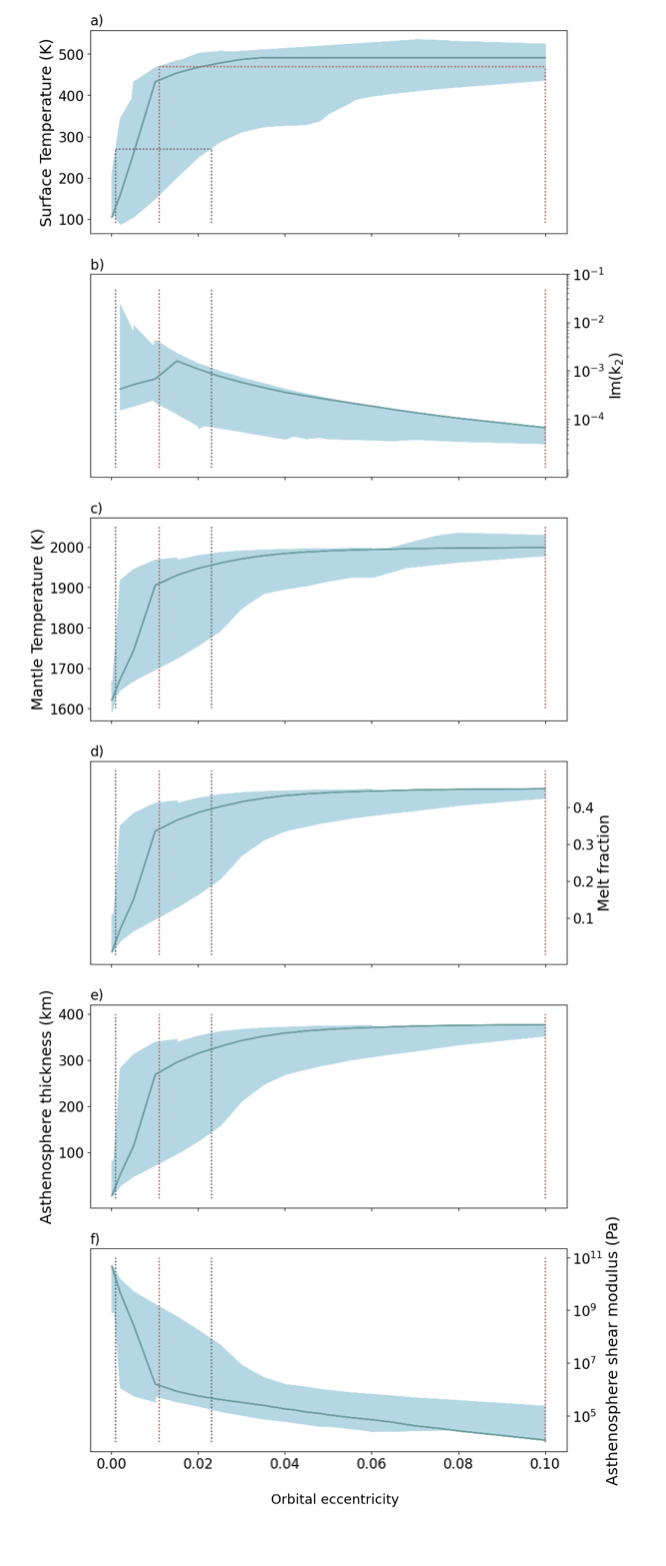}   
    \caption[Sketch]{Computed ranges for the model's output parameters (Table \ref{table:interior parameters} that lead exo-Ios with $a = 4.65$ Roche radii and a radius of $2 \times R_{Io}$ in a stable thermal equilibrium. The blue line corresponds to the Nominal values analysis (Section 5). The grey and red dotted lines show the limits of a 270 K and 470 K respectively surface temperature observation.} \label{fig:finalfig}   
\end{figure*}

As discussed, Figure \ref{fig:finalfig}, which describes the possible constraints we can place on the orbital eccentricity and the interior properties of the moon, assumes a fixed semi-major axis and satellite radius. That means that tidal heating can provide constraints on the interior properties and eccentricity if the orbital period and the size of the satellite are known. To that end, the question of how (and to which accuracy) we can constrain these two parameters in a direct imaging observation appears. 

Even though both \cite{2021AJ....162..181L} and \cite{Mawet2019} find inclinations closer to an edge-on orbit of 78.8 and 89.0 respectively (with large uncertainties), there is no consensus yet of $\epsilon$ Eridani b's inclination.
Nevertheless, we explore possible ways of constraining the properties in scenarios where a system is closer to being edge-on and face-on.
Since the moon is not spatially resolved from the planet, the moon's semi-major axis cannot be constrained in the same way that the planet's one is measured in exoplanet direct imaging observations \citep[e.g., ][]{bohn}.
However, with time series observations of the same target, plausible constraints on the moon's orbital period could be derived.
If the system is edge-on (assuming the planet's and moon's orbits are coplanar), the moon would transit the planet, so the task of deriving the moon's orbital period would be straightforward.
For a face-on system, one could constrain it via the periodicity of the signal coming from patterns on the surface, as hotspots are expected to be present on THEMs' surfaces.
Time series observations of an edge-on system can also infer the satellite's radius \citep{cabrera_schneider_2007}.
However, the size of the satellite can also be deducted via integrating the flux over the observed wavelengths, since the temperature of the body (and the distance) are known.

\subsection{Larger moons}\label{sec:largermoons}

Even though exomoons could be of various sizes and some even more massive than the ones seen in our solar system \citep{Canup2006}, the discussed results in this work refer only to $2 R_{Io}$ exomoons.
Since tidal dissipation depends on the moon's radius we would expect larger satellites to reach warmer thermal equilibrium states, especially in interior regimes with more efficient heat transport than heat piping.
However, larger satellites could be massive enough to sustain a substantial atmosphere \citep{Lammer_2014}.
In this case, direct detection of the thermal emission would be hindered, since the outer atmospheric layer would be relatively colder and detection via spectroscopic signatures would be more promising \citep{Oza_2019}. 
Nonetheless, we point out that a substantial atmosphere would redistribute the tidal energy, making the SED resemble a blackbody when hotspots on the surface shift the SED in bluer wavelenghts \citep{Limbach_2013}.

\subsection{Future model validation} \label{sec:validation}
In this work we applied models previously used to reproduce Io's averaged heat flux \citep{Moore_2003, rovira_2021} to investigate the parameter space, the resulting mean surface temperatures and the possible detection of THEMs via direct imaging. These models have not been yet validated for larger bodies than Io, thus we explore possible observations and complementary models that could in the future validate our results' applicability to analysis of observations.

The observables that we propose are mostly for quantities that are related to tidally induced active volcanism. One notable option is spectral signatures during volcanic outgassing or hotspots. \cite{Oza_2019} introduced a proxy between the amount of outgassed material, the size and the tidal heating rate of the body. This means that we can constrain the tidal heating rate via quantifying volcanically related gases (e.g., sodium, potassium) with spectroscopy, given the size of the body. This measurement could serve as an independent way to deduce the tidal heating rate and thus, the $\text{Im}(k_2)$ of the moon via Equation \ref{eq:tidalflux}. 

Hotspots induced by tidal heating in THEMs could also serve as a way to observe surface inhomogeneities. This has already been applied in exoplanets via single band photometry \citep{Cowan2018,Majeau2012}. Since different regions of the moon's surface would be visible at different orbital phases, spatial inhomogeneities on the surface could be inferred. These variations depend on interior properties in tidally active bodies \citep{STEINKE2020113299}. Thus, one could determine whether possible inferred inhomogeneities are compatible with the amount of tidal heating induced and the interior properties assumed. In addition, the presence of hotspots would be a possible way in which to constrain the moon's orbit period (see Section \ref{section: semimajoraxisconstr}).

Another possible consideration would be validating the calculated $\text{Im}(k_2)$ through orbital constraints. In other words, quantifying whether the $\text{Im}(k_2)$ is compatible with the orbital properties of the putative detection. The orbital circularization timescale depends on the $\text{Im}(k_2)$ of the moon \citep{rovira_2021}. The eccentricity constraints that our analysis provides and the calculated $\text{Im}(k_2)$ should be compatible with the system's age. Such an analysis would require different scenarios to be evaluated, including the possibility for a resonance lock \citep{fuller2016} with the planet's interior, as is likely at present in the case for Saturn's satellite system \citep{LaineyEtAl2020}.

Finally, these considerations are not limited to moons, as exploring tidal heating in exoplanets can serve as a way to better understand the mechanism \citep{Henning_2009, Renaud_Henning_2018}. Applying the discussed model in exoplanets and validating the results would offer a bigger observational sample to test the model's limits.

\section{Conclusions}
We used tidal and thermal models that describe Io's observed surface heat flux \citep{rovira_2021} to investigate Tidally Heated Exomoons (THEM) observability around $\epsilon$ Eridani b. We have concluded that for orbital properties of the same order as solar system satellites (semi major axis of 5.5 Roche radii and eccentricity of 0.009), a $2R_{Io}$ THEM is detectable around $\epsilon$ Eridani b with MIRI. When we take into account uncertainties in the interior properties, $\approx$$40\%$ of such simulated THEMs reach a surface temperature high enough to be detectable. However, $100\%$ of our simulations lead to an observable Super-Io for an eccentricity of 0.02 and a semi-major axis of $\approx$4 Roche-radii. If no significant atmosphere is present, larger moons would experience more intense tidal heating, and they would, thus, also be detectable. 

With a successful thermal flux detection, the models we have presented could place constraints on orbital parameters, with the assumption that the satellites are in thermal equilibrium. If the satellite's semi major axis and radius are inferred (e.g., via exomoon transits) we can place more strict constraints on the orbital eccentricity and additionally on the interior properties, such a melt fraction and thickness of the asthenosphere (see also Table \ref{table:interior parameters}) or whether the moon is in a magma ocean state. For instance, for THEM detection with a surface temperature of 270K and a semi major axis of 4.65 Roche radii an exo-Io would need to have an eccentricity smaller than 0.022 to be in thermal equilibrium. This constraint is looser for higher surface temperatures. In addition, only bodies with melt fractions larger than 0.1 can explain an observation of 470K for this semi-major axis. A number of complementary observations of exomoons and their environment, such as hotspot properties, volcanic outgassing and orbital configuration, could be used to test the applicability of our models after the detection of a THEM.

\begin{acknowledgements}
      This research has been supported by the PEPSci Programme (Planetary and ExoPlanetary Science Programme), NWO, the Netherlands.
\end{acknowledgements}

\bibliography{source}

\end{document}